\documentclass[ALICE,manyauthors]{cernphprep}
\usepackage[comma,square,numbers,sort&compress]{natbib}

\usepackage{lineno}
\modulolinenumbers[5]

\usepackage{xspace}
\usepackage{hyperref}
\usepackage{bm}

\usepackage{color}

\usepackage{orcidlink}
\begin{document}
%

\newcommand{\pp}           {pp\xspace}
\newcommand{\ppbar}        {\mbox{$\mathrm {p\overline{p}}$}\xspace}
\newcommand{\XeXe}         {\mbox{Xe--Xe}\xspace}
\newcommand{\PbPb}         {\mbox{Pb--Pb}\xspace}
\newcommand{\pA}           {\mbox{pA}\xspace}
\newcommand{\pPb}          {\mbox{p--Pb}\xspace}
\newcommand{\AuAu}         {\mbox{Au--Au}\xspace}
\newcommand{\dAu}          {\mbox{d--Au}\xspace}

\newcommand{\s}            {\ensuremath{\sqrt{s}}\xspace}
\newcommand{\snn}          {\ensuremath{\sqrt{s_{\mathrm{NN}}}}\xspace}
\newcommand{\pt}           {\ensuremath{p_{\rm T}}\xspace}
\newcommand{\meanpt}       {$\langle p_{\mathrm{T}}\rangle$\xspace}
\newcommand{\ycms}         {\ensuremath{y_{\rm CMS}}\xspace}
\newcommand{\ylab}         {\ensuremath{y_{\rm lab}}\xspace}
\newcommand{\etarange}[1]  {\mbox{$\left | \eta \right |~<~#1$}}
\newcommand{\yrange}[1]    {\mbox{$\left | y \right |~<~#1$}}
\newcommand{\dndy}         {\ensuremath{\mathrm{d}N_\mathrm{ch}/\mathrm{d}y}\xspace}
\newcommand{\dndeta}       {\ensuremath{\mathrm{d}N_\mathrm{ch}/\mathrm{d}\eta}\xspace}
\newcommand{\avdndeta}     {\ensuremath{\langle\dndeta\rangle}\xspace}
\newcommand{\dNdy}         {\ensuremath{\mathrm{d}N_\mathrm{ch}/\mathrm{d}y}\xspace}
\newcommand{\Npart}        {\ensuremath{N_\mathrm{part}}\xspace}
\newcommand{\Ncoll}        {\ensuremath{N_\mathrm{coll}}\xspace}
\newcommand{\dEdx}         {\ensuremath{\textrm{d}E/\textrm{d}x}\xspace}
\newcommand{\RpPb}         {\ensuremath{R_{\rm pPb}}\xspace}

\newcommand{\nineH}        {$\sqrt{s}~=~0.9$~Te\kern-.1emV\xspace}
\newcommand{\seven}        {$\sqrt{s}~=~7$~Te\kern-.1emV\xspace}
\newcommand{\twoH}         {$\sqrt{s}~=~0.2$~Te\kern-.1emV\xspace}
\newcommand{\twosevensix}  {$\sqrt{s}~=~2.76$~Te\kern-.1emV\xspace}
\newcommand{\five}         {$\sqrt{s}~=~5.02$~Te\kern-.1emV\xspace}
\newcommand{\twosevensixnn}{$\sqrt{s_{\mathrm{NN}}}~=~2.76$~Te\kern-.1emV\xspace}
\newcommand{\fivenn}       {$\sqrt{s_{\mathrm{NN}}}~=~5.02$~Te\kern-.1emV\xspace}
\newcommand{\LT}           {L{\'e}vy-Tsallis\xspace}
\newcommand{\GeVc}         {Ge\kern-.1emV/$c$\xspace}
\newcommand{\MeVc}         {Me\kern-.1emV/$c$\xspace}
\newcommand{\TeV}          {Te\kern-.1emV\xspace}
\newcommand{\GeV}          {Ge\kern-.1emV\xspace}
\newcommand{\MeV}          {Me\kern-.1emV\xspace}
\newcommand{\GeVmass}      {Ge\kern-.2emV/$c^2$\xspace}
\newcommand{\MeVmass}      {Me\kern-.2emV/$c^2$\xspace}
\newcommand{\lumi}         {\ensuremath{\mathcal{L}}\xspace}

\newcommand{\ITS}          {\rm{ITS}\xspace}
\newcommand{\TOF}          {\rm{TOF}\xspace}
\newcommand{\ZDC}          {\rm{ZDC}\xspace}
\newcommand{\ZDCs}         {\rm{ZDCs}\xspace}
\newcommand{\ZNA}          {\rm{ZNA}\xspace}
\newcommand{\ZNC}          {\rm{ZNC}\xspace}
\newcommand{\SPD}          {\rm{SPD}\xspace}
\newcommand{\SDD}          {\rm{SDD}\xspace}
\newcommand{\SSD}          {\rm{SSD}\xspace}
\newcommand{\TPC}          {\rm{TPC}\xspace}
\newcommand{\TRD}          {\rm{TRD}\xspace}
\newcommand{\VZERO}        {\rm{V0}\xspace}
\newcommand{\VZEROA}       {\rm{V0A}\xspace}
\newcommand{\VZEROC}       {\rm{V0C}\xspace}
\newcommand{\Vdecay} 	   {\ensuremath{V^{0}}\xspace}

\newcommand{\ee}           {\ensuremath{e^{+}e^{-}}} 
\newcommand{\pip}          {\ensuremath{\pi^{+}}\xspace}
\newcommand{\pim}          {\ensuremath{\pi^{-}}\xspace}
\newcommand{\kap}          {\ensuremath{\rm{K}^{+}}\xspace}
\newcommand{\kam}          {\ensuremath{\rm{K}^{-}}\xspace}
\newcommand{\pbar}         {\ensuremath{\rm\overline{p}}\xspace}
\newcommand{\kzero}        {\ensuremath{{\rm K}^{0}_{\rm{S}}}\xspace}
\newcommand{\lmb}          {\ensuremath{\Lambda}\xspace}
\newcommand{\almb}         {\ensuremath{\overline{\Lambda}}\xspace}
\newcommand{\Om}           {\ensuremath{\Omega^-}\xspace}
\newcommand{\Mo}           {\ensuremath{\overline{\Omega}^+}\xspace}
\newcommand{\X}            {\ensuremath{\Xi^-}\xspace}
\newcommand{\Ix}           {\ensuremath{\overline{\Xi}^+}\xspace}
\newcommand{\Xis}          {\ensuremath{\Xi^{\pm}}\xspace}
\newcommand{\Oms}          {\ensuremath{\Omega^{\pm}}\xspace}
\newcommand{\degree}       {\ensuremath{^{\rm o}}\xspace}

\newcommand{\Jpsi}         {\ensuremath{\rm{J/\psi}}\xspace}
\newcommand{\Noon}         {\textbf{n$\mathbf{_O^O}$n}}
\newcommand{\Wgp}{\ensuremath{W_{\gamma \mathrm{Pb}}}\xspace}
\newcommand{\ptdimuon}     {\ensuremath{p^{\mu\mu}_{\rm T}}\xspace}
\newcommand{\mdimuon}     {\ensuremath{m_{\mu\mu}}\xspace}
\newcommand{\pttwo}        {\ensuremath{p^{2}_{\rm T}}\xspace}
\newcommand{\mant}         {\ensuremath{|t|}\xspace}

\newcommand{\BR}           {\ensuremath{\rm BR}\xspace}
\newcommand{\axe}          {\ensuremath{\rm{Acc\times\epsilon}}\xspace}
\newcommand{\fC}           {\ensuremath{f_{\rm C}}\xspace}
\newcommand{\fD}           {\ensuremath{f_{\rm D}}\xspace}

\begin{titlepage}
\PHyear{2023}       
\PHnumber{080}      
\PHdate{04 May}  

\title{First measurement of the $|\mathbf{\textit t}|$-dependence of incoherent $\mathbf{\Jpsi}$ photonuclear production}
\ShortTitle{$\mant$-dependence of incoherent $\Jpsi$ photonuclear production}   

\Collaboration{ALICE Collaboration\thanks{See Appendix~\ref{app:collab} for the list of collaboration members}}
\ShortAuthor{ALICE Collaboration} 

\begin{abstract}
The first measurement of the cross section for incoherent photonuclear production of $\Jpsi$ vector mesons as a function of the Mandelstam $\mant$ variable is presented. The measurement was carried out with the ALICE detector at midrapidity, $|y|<0.8$, using ultra-peripheral collisions of Pb nuclei at a centre-of-mass energy per nucleon pair of $\sqrt{s_{\mathrm{NN}}}~=~5.02$~TeV. This rapidity interval corresponds to a Bjorken-$x$ range $(0.3$--$1.4)\times 10^{-3}$. Cross sections are given in five $\mant$ intervals in the range $0.04<\mant<1$~GeV$^2$ and compared to the predictions by different models. Models that ignore quantum fluctuations of the gluon density in the colliding hadron predict a $\mant$-dependence of the cross section much steeper than in data. The inclusion of such fluctuations in the same models provides a better description of the data. 
\end{abstract}

\end{titlepage}

\setcounter{page}{2} 


The fundamental structure of protons, neutrons, and nuclei is described in terms of quarks and gluons by quantum chromodynamics (QCD). 
A new phenomenon called gluon saturation---a dynamic equilibrium between the production and annihilation of gluons---is predicted by QCD~\cite{Gribov:1983ivg}. While the high-energy limit of QCD has been found to be dominated by the gluon contribution in proton targets~\cite{H1:2015ubc}, experimental work is yet needed to determine the onset of gluon saturation~\cite{Morreale:2021pnn}. Besides protons at high energy, saturation is expected for large nuclei at even lower energies~\cite{McLerran:1993ni}, thus the study of the structure of heavy ions is an attractive area of exploration within the current collider experiments. The search for the onset of saturation has motivated the construction of dedicated QCD facilities such as the future Electron--Ion Collider~\cite{AbdulKhalek:2021gbh}.

Photons are ideal probes to study the interior of nuclei. In this context, the diffractive photoproduction of a vector meson, like the $\Jpsi$, is of particular interest because of its sensitivity to both the average and the variance of spatial distribution of the gluon field inside nuclei~\cite{Klein:2019qfb}. In this process, a quasi-real photon emitted by one of the highly Lorentz-contracted nuclei interacts via the exchange of at least two gluons with the other nucleus, producing the vector meson~\cite{Ryskin:1992ui}. 

This process can be divided in two contributions: coherent and incoherent production. The former refers to photon interactions with the colour field of the whole nucleus, and the latter to photon interactions with only one nucleon inside the nucleus. The incoherent production can be further divided in the interaction with a full nucleon or the interaction with sub-nucleon sized structures inside the nucleon; the latter is known as the dissociative contribution. The square of the momentum transferred during the interaction, the Mandelstam variable $\mant$, is related through a Fourier transform to the distribution of nuclear matter in the impact-parameter plane. This implies that collisions with a large scattering object, such as the whole nucleus, occur at small $\mant$, which for the case of Pb ions means $\mant\lesssim 0.01$~GeV$^2$. In the same way, collisions with a small object, like a nucleon, lead to larger $\mant$ values of the order of 0.1~GeV$^2$. If there are collisions with even smaller objects at a sub-nucleon scale, they would have even larger $\mant$. In the Good--Walker approach~\cite{Good:1960ba}, the coherent process is related to the average spatial distribution of gluons in the transverse plane, and the incoherent case is related to its variance~\cite{Miettinen:1978jb}. The applicability of this approach to LHC data may have some caveats as discussed in~\cite{Klein:2023zlf}. A recent study using this approach~\cite{Mantysaari:2016ykx} demonstrated the importance of including fluctuations of spatial distributions of gluons to describe the $\mant$-dependence of the dissociative cross section off protons measured at HERA~\cite{H1:2013okq}. Further work in this direction~\cite{Cepila:2016uku} revealed that the energy dependence of the dissociative process provides another signature for saturation. When the gluon saturation regime is reached, all gluon configurations in the proton appear similar, thus the cross section, which is proportional to the variance of the gluon field, decreases as the energy increases. Note that larger values of \mant are expected to be more sensitive to fluctuations, thus it is important to study the energy dependence at different values of \mant, where a decrease of the cross section with increasing energy would be a signature of saturation.

Although the dissociative production of $\Jpsi$ off protons has been measured at HERA~\cite{H1:2013okq}, until now this process has not been measured using heavy-ion targets. Most of the experimental effort has been put on coherent vector meson photoproduction. At high energies, this has been carried out using photon-induced processes in ultra-peripheral heavy-ion collisions (UPCs) at the Large Hadron Collider (LHC)~\cite{Baltz:2007kq,Contreras:2015dqa,Klein:2019qfb}. The diffractive photoproduction of a $\Jpsi$ vector meson at the LHC has a very clean experimental signal with a sizeable cross section. The coherent photoproduction of a $\Jpsi$ off the Pb nuclei has been measured at the LHC at two different centre-of-mass energies per nucleon pair, \twosevensixnn and 5.02~TeV, by the ALICE~\cite{ALICE:2013wjo,ALICE:2019tqa,ALICE:2021gpt}, CMS~\cite{CMS:2016itn}, and~LHCb~\cite{LHCb:2021bfl} Collaborations. Together, these measurements cover a range in $\Jpsi$ rapidity of $|y|<4.5$. More recently, the ALICE Collaboration performed the first measurement of the $\mant$-dependence of the coherent $\Jpsi$ photoproduction cross section~\cite{ALICE:2021tyx}, and the STAR Collaboration studied the structure of the deuteron through the $\mant$-dependence of \Jpsi diffractive photoproduction in deuteron--gold collisions~\cite{STAR:2021wwq}. The ALICE Collaboration also presented a measurement of the cross section for incoherent $\Jpsi$ production at midrapidity~\cite{ALICE:2013wjo}.

In recent years, great theoretical interest has been given to incoherent $\Jpsi$ photoproduction~\cite{Cepila:2018zky,Luszczak:2017dwf,Goncalves:2020vdp,Sambasivam:2019gdd}, and in particular to its $\mant$-dependence~\cite{Mantysaari:2017dwh,Cepila:2017nef,Guzey:2018tlk}. Theoretical approaches that describe correctly the coherent production process differ widely in their predictions for incoherent production, which is particularly sensitive to spatial fluctuations of sub-nucleon degrees of freedom. Note that a better assessment of such quantum fluctuations would significantly improve the determination of the initial stage of nuclear collisions at high energies~\cite{Mantysaari:2020axf}. 

In this Letter, the first measurement of the $\mant$-dependence of the  incoherent photonuclear production of a $\Jpsi$ vector meson is presented. The measurement was carried out in the rapidity range $|y|<0.8$ using UPCs of Pb nuclei at \fivenn. Cross sections are presented in five $\mant$ intervals in the range $0.04<\mant<1$ GeV$^2$. The measurement is compared to the predictions of the models discussed later on, finding that the contribution of fluctuations at a sub-nucleon scale is important to describe the data. 

This analysis is based on the data set collected during the 2018 \PbPb data-taking period. It utilises the same trigger and follows the same analysis strategy as in Ref.~\cite{ALICE:2021tyx}. The luminosity of the analysed sample is $(232 \pm 7)~\mu$b$^{-1}$. The measured $\Jpsi$ mesons have a rapidity $|y|<0.8$, corresponding to Bjorken-$x$ values within $(0.3$--$1.4)\times 10^{-3}$, and transverse momentum $0.2<\pt<1$ GeV$/c$. Owing to the small virtuality of the quasi-real photons, in the kinematic region studied here $\mant = \pt^2$. According to the STARlight Monte Carlo~\cite{Klein:2016yzr}, the difference between the mean \mant and $\pt^2$ in each interval is less than 0.4\%. As \pt is conjugate to impact parameter, which in UPC is large, interference effects are important only at $\pt$ below 10~MeV/$c$ and are negligible for the $\pt$ range of this measurement~\cite{Klein:1999gv}.

The $\Jpsi$ was reconstructed using its decay into a $\mu^+\mu^-$ pair. The signature of these events is then two tracks in an otherwise empty  detector. The only other particles that may be present in such an event are the products from the dissociation of the interacting nucleus; these particles would appear near beam rapidities.  The muons were measured with the central barrel detectors of ALICE~\cite{ALICE:2008ngc,ALICE:2014sbx}: the ALICE Inner Tracking System (ITS)~\cite{ALICE:2010tia} and the Time Projection Chamber (TPC)~\cite{Alme:2010ke}, both of them covering the full azimuthal angle and surrounded by a large solenoid magnet producing a magnetic field of 0.5~T. Any other activity in the event was vetoed by the V0~\cite{ALICE:2013axi} and the AD~\cite{Broz:2020ejr}, which are scintillator based detectors consisting of two arms each, located at both sides of the nominal interaction point along the beam axis. They cover the pseudorapidity ranges $2.8 < \eta < 5.1$ and $-3.7 < \eta < -1.7$ (V0), and $4.8 < \eta < 6.3$ and ${-7.0 < \eta < -4.9}$ (AD). Each arm of V0 and AD has a time resolution smaller than 1 ns. 

The tracks were required to have opposite electric charges and to leave signals in both the ITS and the TPC. Their pseudorapidity was constrained to $|\eta|<0.8$ in order to have a large reconstruction efficiency. The muons were identified by requiring an ionisation energy loss, measured in the TPC, compatible with the muon hypothesis. For the momentum range of the muons in this analysis (0.5 to 3~GeV/$c$) this criterion rejects completely the contribution from the electron decay channel. The two tracks were required to form a common interaction vertex with a coordinate along the nominal beam line $|z_{\rm vtx}|<10$~cm to have uniform acceptance. 

\begin{figure}[!t]
\centering
\includegraphics[width=0.48\textwidth]{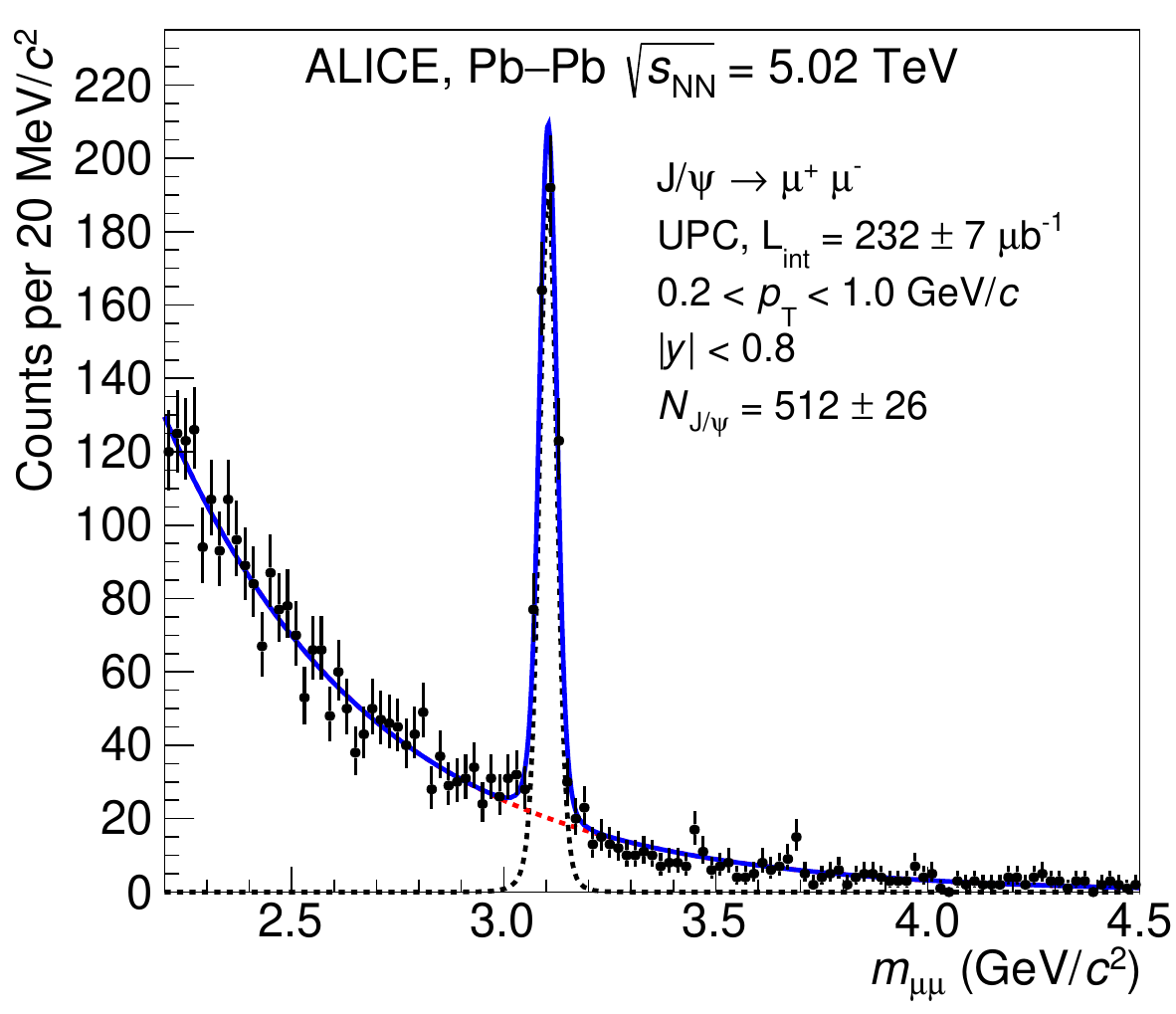}
\includegraphics[width=0.48\textwidth]{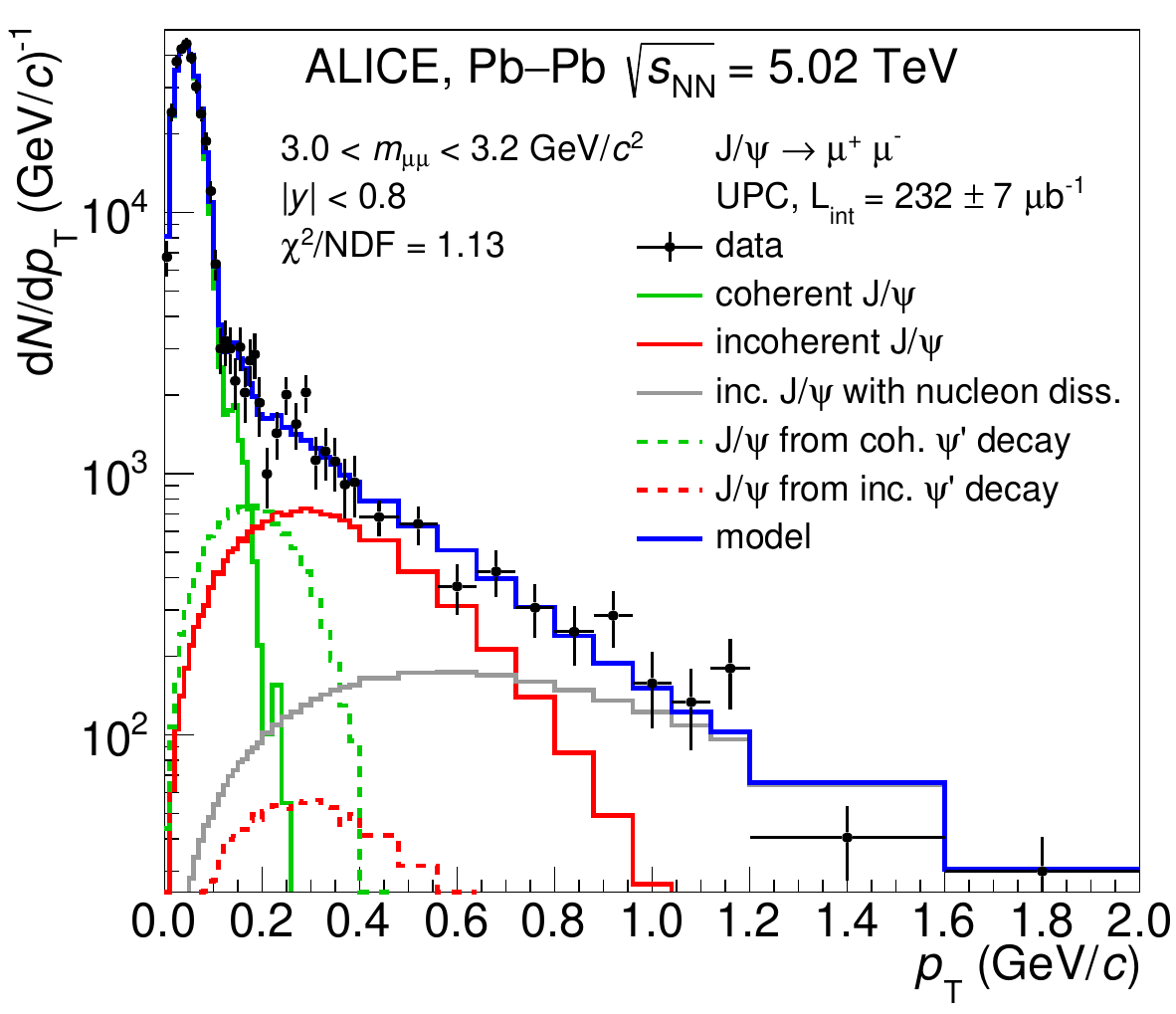}
\caption{\label{fig:fits} 
Left: Invariant mass distribution of muon pairs (full symbols) and fit to a model (solid blue line, see text). Right: transverse momentum distribution of muon pairs with $3.0 < m_{\mu\mu} < 3.2$~GeV/$c^2$ (full symbols) and fit to a model (solid blue line) along with the different contributions to the fit (other lines, see text).}
\end{figure}

The $\Jpsi$ yield, $N_{\Jpsi}$, was extracted by fitting the muon-pair invariant-mass ($m_{\mu\mu}$) distribution  with two contributions: a double sided Crystal Ball distribution~\cite{ALICE:2015ikg} to represent the signal and an exponential to describe the background. An unbinned extended likelihood fit was performed in each one of the five $\mant$ intervals. The left panel of Fig.~\ref{fig:fits} shows the fit to the total sample. The extracted $\Jpsi$ yield is $512\pm 26$~(stat.). This yield is dominated by the contribution of incoherent processes, but it still has a remaining background that has to be subtracted. The amount of background is obtained by analyzing the transverse momentum distribution.

The $\Jpsi$ yield originates from three contributions: coherent and incoherent production, as well as feed-down from $\psi^\prime$ diffractive photoproduction. The background to the yield from incoherent production, $N^{\rm inc}_{\Jpsi}$, was subtracted in each $\mant$ range using the ratio of the number of $\Jpsi$ from coherent (feed-down) to incoherent production \fC (\fD) such that $N^{\rm inc}_{\Jpsi} = N_{\Jpsi}/(1+\fC+\fD$).

The \fC and \fD ratios were determined from a binned extended likelihood fit to the transverse-momentum distribution of the $\Jpsi$ yield in the range ${3.0<m_{\mu\mu}< 3.2}$~GeV$/c^2$. The $\Jpsi$ yields were obtained by performing in each bin a fit to the invariant mass distribution, using the model described above. The fit to the transverse-momentum distribution is shown in the right panel of Fig.~\ref{fig:fits}. The data were fitted to the sum of five templates. Four of them, describing the contributions of coherent and incoherent production of both \Jpsi and $\psi^\prime$, are obtained with the STARlight Monte Carlo. The charmonium states are assumed to be transversely polarised as expected for photoproduction processes~\cite{ALICE:2023svb}. The shape of the transverse momentum distribution is given by the target form factor which in turn is obtained by the Fourier transform of the target profile in the impact-parameter space. This presents the physics modelling implemented in STARlight. It is known that STARlight does not describe correctly the shape of coherent $\Jpsi$ production in the range $\pt < 0.11$ GeV/$c$~\cite{ALICE:2021tyx} when using the default value of the parameters for the nuclear form factor. At the same time, the data can be described using a different value that was found by reweighting the STARlight templates. The template corresponding to coherent \Jpsi production is the only one affected by such a procedure. Moreover, whereas the effect of the reweighting is important for $\pt<0.2$~GeV/$c$, which is outside the kinematic region of the measurement presented here and contains about 99\% of the coherent cross section, it corresponds to no more than a 2\% modification of the final incoherent \Jpsi cross section in the lowest \mant range. Note that the STARlight implementation of the incoherent process does not include the dissociative contribution. For this reason a fifth template was added, which uses the H1 parameterisation of dissociative production off protons~\cite{H1:2013okq}. In this parametrisation, the values corresponding to the H1 high-energy sample were used. Although the parameters were obtained for free protons, they describe well the shape of the distribution, as shown in Fig.~\ref{fig:fits}. The measured ratio $R$ of the coherent $\psi'$ to $\Jpsi$ cross sections~\cite{ALICE:2021gpt} fixes the normalisation of the $\psi^\prime$ templates; here the acceptance and efficiency of each decay channel is taken into account. This leaves the normalisations of the templates describing coherent, incoherent and dissociative $\Jpsi$ photoproduction as free parameters. The value of $R$ was assumed to be the same for the ratio of incoherent cross sections, which has not yet been measured in heavy-ion collisions, but was measured at HERA by the H1~\cite{H1:2002yab} and ZEUS~\cite{ZEUS:2022sxn} Collaborations in electron--proton collisions and found in agreement with the value of $R$ measured by ALICE~\cite{ALICE:2021gpt}  for coherent \Jpsi production in Pb--Pb collisions. The $\chi^2$ per degree of freedom of the fit is 1.13. The values for $f_{\rm C}$ and $f_{\rm D}$ are listed in Table~\ref{tab:results}.

The photonuclear cross section in each $\mant$ interval was computed as
\begin{equation}
\frac{\text{d}\sigma_{\gamma{\rm Pb}}}{\text{d}\mant} = \frac{1}{2n_{\gamma{\rm Pb}}}\frac{N^\text{inc}_{\Jpsi}}{(\text{Acc}\times\varepsilon)^\text{inc}_{\Jpsi} \times \text{BR}({\Jpsi} \rightarrow \mu^+\mu^-) \times \mathcal{L} \times \Delta y \times \Delta\mant} \, ,
\label{eq:IncCrossSection}
\end{equation}
where $n_{\gamma{\rm Pb}}=84.9 \pm 1.7$ is the photon flux at $y=0$, obtained in the semiclassical formalism following the prescription detailed in Ref.~\cite{Contreras:2016pkc}; the branching ratio $\text{BR}(\Jpsi \rightarrow \mu^+\mu^-)=(5.961 \pm 0.033)\%$ is from Ref.~\cite{Workman:2022ynf}; the luminosity $\mathcal{L}=(232 \pm 7)~\mu$b$^{-1}$ was determined using reference triggers with cross sections measured in van der Meer scans~\cite{ALICE:2022xir}; and $(\text{Acc}\times\varepsilon)^\text{inc}_{\Jpsi}$ is the acceptance times efficiency. This last term is the product of three contributions. The first one takes into account the response of the detector to the muon tracks; this contribution was obtained from generated STARlight events which were passed through a simulation of the ALICE detector using GEANT~3.21~\cite{Brun:1994aa} and the full analysis chain. As shown in  column $(\text{Acc}\times\varepsilon)_\text{MC}$ of Table~\ref{tab:results}, the correction depends on \mant due to the trigger, which requires tracks that are back-to-back in azimuth~\cite{ALICE:2021tyx}. The second term contributing to $(\text{Acc}\times\varepsilon)^\text{inc}_{\Jpsi}$ accounts for veto inefficiencies due to pile-up of other collisions leaving a signal in AD or V0 and amounts to $0.940\pm0.028$. The third term corrects the yield for events lost because the dissociation of the nucleus produces particles leaving a signal in the AD detectors; it amounts to $0.637\pm0.024$. These last two factors are \mant independent, with the quoted uncertainty originating from the size of the control data samples used to determine them.

\begin{table}
	\caption{Measured cross sections, shown in the last column, and the numerical values used to compute them according to 
	Eq.~(\ref{eq:IncCrossSection}). The uncertainties on $N_{\Jpsi}$ and $(\text{Acc}\times\varepsilon)_\text{MC}$ are statistical; those on $f_\textrm{C}$ and $f_\textrm{D}$ are each correlated systematic; those on the cross sections are (in this order) statistical, uncorrelated systematic, and correlated systematic.}
	\centering
	\begin{tabular}{|c|c|c|c|c|c|c|} 
	\hline
	\mant (GeV$^2$)
	    & $N_{\Jpsi}$
	    & $f_\textrm{C}$ (\%)
	    & $f_\textrm{D}$ (\%)
        & $(\text{Acc}\times\varepsilon)_\text{MC}$ (\%)
        & $\frac{\text{d}\sigma_{\gamma\text{Pb}}}{\text{d}|t|}$ ($\mu$b/GeV$^2$) \\
    \hline\hline
    $(0.040,0.080)$ & $128 \pm 12$  & $9.4 \pm 0.8$     & $81.9 \pm 11.7$ & $3.39 \pm 0.03$ & $21.8 \pm 2.1 \pm 0.3 \pm 2.1$ \\
    $(0.080,0.152)$ & $127 \pm 12$  & $0.024 \pm 0.002$ & $36.0 \pm 4.9$  & $3.03 \pm 0.02$ & $19.1 \pm 1.9 \pm 0.3 \pm 1.5$ \\
    $(0.152,0.258)$ & $85 \pm 10$   & $0$               & $9.3 \pm 1.0$   & $2.49 \pm 0.02$ & $13.1 \pm 1.6 \pm 0.4 \pm 0.9$ \\
    $(0.258,0.477)$ & $86 \pm 11$   & $0$               & $4.9 \pm 0.4$   & $2.04 \pm 0.02$ & $8.1 \pm 1.1 \pm 0.1 \pm 0.6$ \\
    $(0.477,1.000)$ & $86 \pm 11$   & $0$               & $2.7 \pm 0.2$   & $1.57 \pm 0.02$ & $4.6 \pm 0.6 \pm 0.1 \pm 0.3$ \\
    \hline
	\end{tabular}
	\label{tab:results}
\end{table}

The following systematic uncertainties were studied and their effect on the cross section is summarised in Table~\ref{tab:SystematicUncertainties}. To study the stability of the background model, the lower and upper limits to the invariant-mass fits to extract the signal were varied in the range of 2--2.5 and 4--5~GeV/$c^2$, respectively. The values of the tail parameters of the Crystal Ball distribution were also modified; the central values and the variations were obtained by fitting STARlight simulated events. The total effect on the cross section varies in the different $\mant$ ranges between 1\% and 2.9\%. The detector does not have a uniform acceptance for tracks from collisions happening far from the nominal interaction point; to study the quality of the detector description for these extreme cases the selection $|z_{\rm vtx}| < 10$~cm was extended to $|z_{\rm vtx}| < 15$~cm, resulting in uncertainties at the level of up to 2.9\%. There are three contributions to the uncertainties on the \fC and \fD factors: the uncertainties from the fit (driven by statistical fluctuations), the uncertainty from the reweighting procedure, and the effect of varying the value of $R$ within the experimental uncertainties. The uncertainty on \fC is driven by the uncertainty from the fit to the \pt distribution and leads to uncertainties in the measured cross section up to 0.4\%. The uncertainty on \fD is driven by the uncertainty on the measured value of $R$ and produces an effect from 0.2\% to 6.5\%. The uncertainty on the luminosity has two contributions which were added in quadrature: from the measurement of the reference cross sections in van der Meer scans (2.5\%~\cite{ALICE:2022xir}) and from the determination of the live-time of the trigger used in this analysis (1.5\%). The correction for pile-up utilises an independent sample to obtain the dependence of pile-up on the average rate of inelastic scattering; this dependence is linear and the corresponding uncertainty comes from a fit to these data. The effect on the cross section is 3\%. The probability of dissociation products leaving a signal in AD was studied with an independent control sample as a function of the amount of activity around beam rapidity. The propagation of the statistical uncertainty of the correction factors when applied to this sample produces a 3.8\% effect. The uncertainty of 2\% per track on the matching of ITS and TPC track segments was estimated from the difference between matching efficiencies in data and MC simulations. Contributions from both tracks were added in quadrature, giving a total of 2.8\%. The trigger efficiency uncertainty was determined using control data samples and amounts to 1.3\%. The uncertainty on the branching ratio was taken from Ref.~\cite{Workman:2022ynf}. The uncertainty on the photon flux was estimated by varying the nuclear radius parameter of the Woods--Saxon distribution in Pb, used in the Glauber model, according to neutron-skin measurements~\cite{Abrahamyan:2012gp} and amounts to 2\%~\cite{ALICE:2021tyx}. All uncertainties except for the signal extraction and the selection on $|z_\text{vtx}|$ are correlated in $\mant$.

\begin{table}[t!]
	\caption{Summary of the identified systematic uncertainties to the cross section. The numbers in parentheses denote a range of values in the different $\mant$ intervals. Except for the first two uncertainties, all others are correlated in $\mant$.}
	\centering
	\begin{tabular}{|c|c|} 
	\hline
    Source & Uncertainty~(\%) \\
    \hline\hline
    Signal extraction               & $(1.0,2.9)$ \\\hline
    Selection on $|z_\text{vtx}|$   & $(0.0,2.9)$ \\\hline
    $f_{\rm C}$                     & $(0.0,0.4)$ \\\hline
    $f_{\rm D}$                     & $(0.2,6.5)$ \\\hline
    Integrated luminosity                   & 2.9 \\\hline
    Veto inefficiency due to pile-up        & 3.0 \\\hline
    Veto inefficiency due to dissociation   & 3.8 \\\hline
    ITS-TPC tracking                        & 2.8 \\\hline
    Trigger efficiency                      & 1.3 \\\hline
    Branching ratio                         & 0.6 \\\hline
    Photon flux                             & 2.0 \\\hline
	\end{tabular}
	\label{tab:SystematicUncertainties}
\end{table}

\begin{figure}[!t]
\centering
\includegraphics[width=0.75\textwidth]{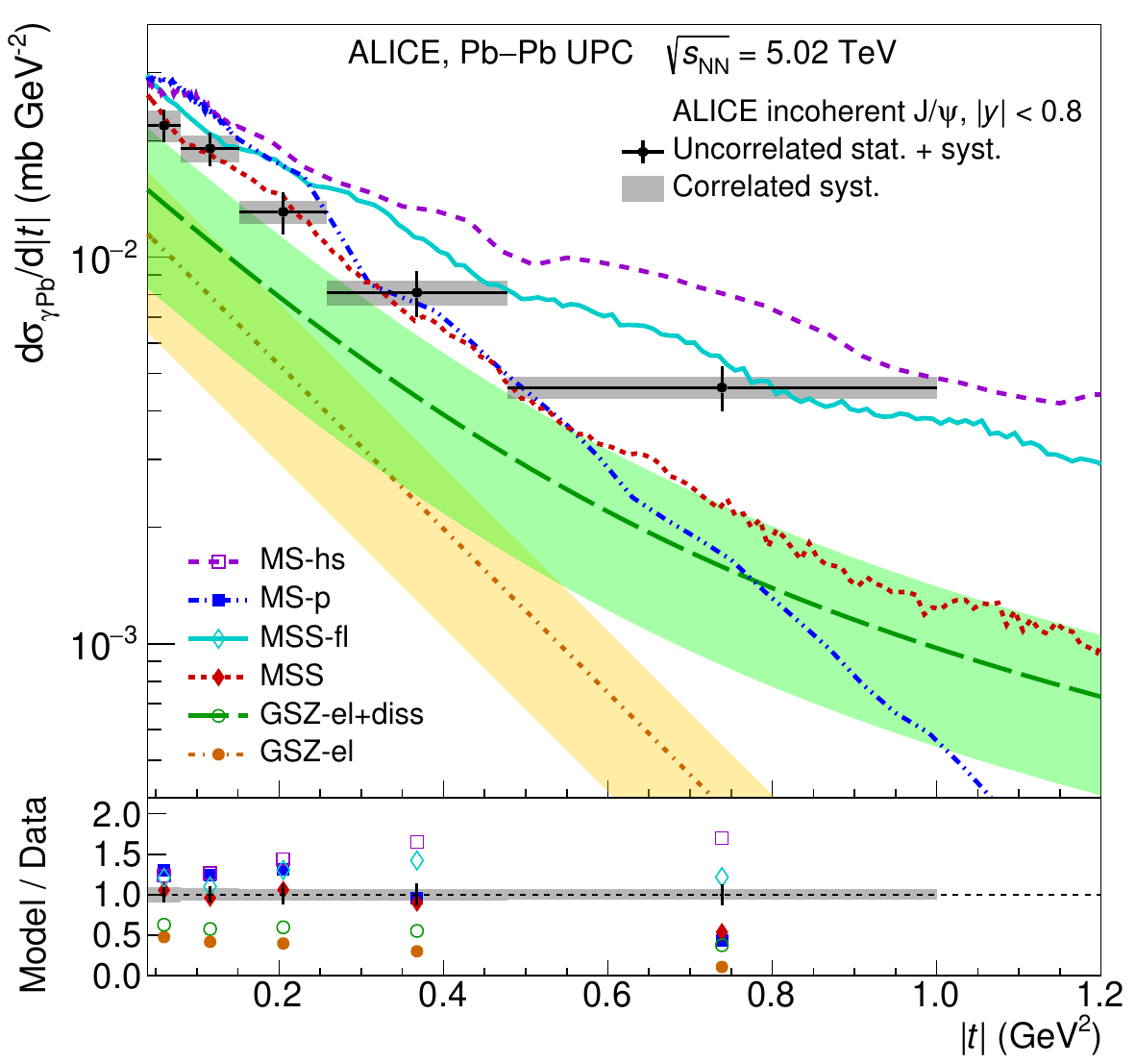}
\caption{Cross section for the incoherent photoproduction of $\Jpsi$ vector mesons in ultra-peripheral Pb--Pb collisions at $\snn= 5.02$~TeV measured at midrapidity. The uncorrelated uncertainty (statistical and systematic added in quadrature) is indicated with the vertical bar, while the correlated uncertainty by the grey band. The width of each $\mant$ range is given by the horizontal bars. The lines show the predictions of the different models described in the text. The bottom panel presents the ratio of the integral of the predicted to that of the measured cross section in each $\mant$ range. The relative uncertainties on the ratios calculated from GSZ are 45\%.
}
\label{fig:crossSection}
\end{figure}

The  cross sections for the incoherent photoproduction of $\Jpsi$ vector mesons in ultra-peripheral Pb--Pb collisions at  $\snn= 5.02$~TeV as a function of $\mant$ measured at midrapidity, $|y|<0.8$, are listed in Table~\ref{tab:results} and depicted in Fig.~\ref{fig:crossSection}. 

The measurements are compared to the work of three groups. Each of them provides two predictions: one including only the elastic interaction with single nucleons, and another where a dissociative-like component is contained. The models are framed in the Good--Walker picture, which naturally considers all possible configurations of the hadron participating in the interaction. 
The model by M\"antysaari and Schenke (MS)~\cite{Mantysaari:2017dwh} includes saturation through the IPSat model~\cite{Kowalski:2003hm} and offers two predictions. In one, sub-nucleon fluctuations are not considered (MS-p), whereas in the other the proton is composed of three hot spots whose positions in the impact-parameter plane change event-by-event and fluctuations in the saturation scale are introduced (MS-hs). A similar model, labeled MSS in Fig. 2, was recently published~\cite{Mantysaari:2022sux}, the main difference in respect of the MS model is that instead of using the IPsat model, it solves the JIMWLK equation (see Refs.~\cite{Marquet:2010cf,Mueller:2001uk}) to incorporate saturation effects. This model also offers two predictions: with (MSS-fl) and without nucleon substructure fluctuations (MSS). The model by Guzey, Strikman, and Zhalov (GSZ)~\cite{Guzey:2018tlk} expresses the incoherent cross section as the sum of an elastic and a dissociative part (GSZ-el+diss), both parameterised from HERA data, multiplied by a common factor representing shadowing---the fact that the gluon distribution in nuclei is not just the sum of gluon distributions in constituent nucleons, see e.g. Ref.~\cite{Armesto:2006ph}---computed within the leading-twist approximation~\cite{Frankfurt:2011cs}. The inclusion of the dissociative component is interpreted by the authors within a Good--Walker approach as due to quantum fluctuations of the target. When the dissociative part is excluded (GSZ-el), the differential cross section is suppressed in the region of larger $\mant$. The uncertainty bands reflect the uncertainties on the parameters of the leading-twist approximation.

When comparing the data with the model predictions, as shown in Fig.~\ref{fig:crossSection}, two aspects should be considered: the normalisation, mainly linked to the scaling from proton to nuclear targets, and the $\mant$-dependence, driven by the size of the scattering object. None of the models describe both aspects of data. With regards to the normalisation, it is worth noting that the same models must also describe the coherent cross section~\cite{ALICE:2021gpt}, hence a global scaling factor, such as what would be obtained by using a different prescription for the wave function~\cite{Cepila:2019skb}, would not necessarily improve the agreement of the model with both the coherent and incoherent cross sections. As for the $\mant$-dependence of the cross section, the predictions of the three theory groups substantially improve after the inclusion of sub-nucleon fluctuations, which modify the $\mant$-dependence by making it less steep. It is interesting to compare the MS-p and MSS predictions. The latter shows a flattening of the spectra at larger $\mant$. It originates from colour charge fluctuations which change the incoherent cross section to a power-law like behaviour  in this region~\cite{Mantysaari:2022sux}. This observation reinforces the importance of quantum fluctuations at large $\mant$.

The cross section integrated over the interval $0.04 < \mant < 1$ GeV$^2$, measured in the rapidity region $|y| < 0.8$, is $\sigma_{\gamma\mathrm{Pb}} = (7.82 \pm 0.39 \pm 0.57)~\mu$b, where the listed uncertainties are statistical and systematic, respectively. The corresponding cross sections, in $\mu$b, for the models are 7.4, 11.8, 6.6, 9.8, $2.3 \pm 1.0$, and $4.1 \pm 1.8$ for MS-p, MS-hs, MSS, MSS-fl, GSZ-el, and GSZ-el+diss, respectively.

In summary, the first measurement of the incoherent photonuclear production of $\Jpsi$ is presented in this Letter. The measurement was carried out at midrapidity, in a range corresponding to Bjorken-$x$ within $(0.3$--$1.4)\times 10^{-3}$, in Pb--Pb UPCs at \fivenn. Cross sections for five ranges in $\mant$ within $0.04<\mant<1$ GeV$^2$ are reported. None of the models describes both the absolute normalisation and the $\mant$-dependence observed in the data. However, a reasonably good description of the measured \mant-slope is achieved when the predicted dependence is softened by the inclusion of scattering structures at a sub-nucleon scale. These results confirm the importance of sub-nucleon fluctuations to describe the measured incoherent \Jpsi process at high energies, representing the first experimental step to use the quantum fluctuations of the gluon field to search for saturation effects in heavy nuclei. In addition, this measurement, when confronted to models, demonstrates that the contribution of the dissociative component to the total incoherent cross section depends on \mant. Thus, future analyses shall study the incoherent production of \Jpsi as a function of rapidity and \mant~\cite{Guzey:2013jaa}. Finally, this analysis, together with recent measurements~\cite{ALICE:2019tqa,CMS:2016itn}, indicate that new or improved theoretical models are needed to describe simultaneously the energy and $\mant$-dependence of both the coherent and the incoherent processes of $\Jpsi$ photoproduction, to gain a better understanding of saturation effects at a more fundamental level.


\newenvironment{acknowledgement}{\relax}{\relax}
\begin{acknowledgement}
\section*{Acknowledgements}

The ALICE Collaboration would like to thank all its engineers and technicians for their invaluable contributions to the construction of the experiment and the CERN accelerator teams for the outstanding performance of the LHC complex.
The ALICE Collaboration gratefully acknowledges the resources and support provided by all Grid centres and the Worldwide LHC Computing Grid (WLCG) collaboration.
The ALICE Collaboration acknowledges the following funding agencies for their support in building and running the ALICE detector:
A. I. Alikhanyan National Science Laboratory (Yerevan Physics Institute) Foundation (ANSL), State Committee of Science and World Federation of Scientists (WFS), Armenia;
Austrian Academy of Sciences, Austrian Science Fund (FWF): [M 2467-N36] and Nationalstiftung f\"{u}r Forschung, Technologie und Entwicklung, Austria;
Ministry of Communications and High Technologies, National Nuclear Research Center, Azerbaijan;
Conselho Nacional de Desenvolvimento Cient\'{\i}fico e Tecnol\'{o}gico (CNPq), Financiadora de Estudos e Projetos (Finep), Funda\c{c}\~{a}o de Amparo \`{a} Pesquisa do Estado de S\~{a}o Paulo (FAPESP) and Universidade Federal do Rio Grande do Sul (UFRGS), Brazil;
Bulgarian Ministry of Education and Science, within the National Roadmap for Research Infrastructures 2020-2027 (object CERN), Bulgaria;
Ministry of Education of China (MOEC) , Ministry of Science \& Technology of China (MSTC) and National Natural Science Foundation of China (NSFC), China;
Ministry of Science and Education and Croatian Science Foundation, Croatia;
Centro de Aplicaciones Tecnol\'{o}gicas y Desarrollo Nuclear (CEADEN), Cubaenerg\'{\i}a, Cuba;
Ministry of Education, Youth and Sports of the Czech Republic, Czech Republic;
The Danish Council for Independent Research | Natural Sciences, the VILLUM FONDEN and Danish National Research Foundation (DNRF), Denmark;
Helsinki Institute of Physics (HIP), Finland;
Commissariat \`{a} l'Energie Atomique (CEA) and Institut National de Physique Nucl\'{e}aire et de Physique des Particules (IN2P3) and Centre National de la Recherche Scientifique (CNRS), France;
Bundesministerium f\"{u}r Bildung und Forschung (BMBF) and GSI Helmholtzzentrum f\"{u}r Schwerionenforschung GmbH, Germany;
General Secretariat for Research and Technology, Ministry of Education, Research and Religions, Greece;
National Research, Development and Innovation Office, Hungary;
Department of Atomic Energy Government of India (DAE), Department of Science and Technology, Government of India (DST), University Grants Commission, Government of India (UGC) and Council of Scientific and Industrial Research (CSIR), India;
National Research and Innovation Agency - BRIN, Indonesia;
Istituto Nazionale di Fisica Nucleare (INFN), Italy;
Japanese Ministry of Education, Culture, Sports, Science and Technology (MEXT) and Japan Society for the Promotion of Science (JSPS) KAKENHI, Japan;
Consejo Nacional de Ciencia (CONACYT) y Tecnolog\'{i}a, through Fondo de Cooperaci\'{o}n Internacional en Ciencia y Tecnolog\'{i}a (FONCICYT) and Direcci\'{o}n General de Asuntos del Personal Academico (DGAPA), Mexico;
Nederlandse Organisatie voor Wetenschappelijk Onderzoek (NWO), Netherlands;
The Research Council of Norway, Norway;
Commission on Science and Technology for Sustainable Development in the South (COMSATS), Pakistan;
Pontificia Universidad Cat\'{o}lica del Per\'{u}, Peru;
Ministry of Education and Science, National Science Centre and WUT ID-UB, Poland;
Korea Institute of Science and Technology Information and National Research Foundation of Korea (NRF), Republic of Korea;
Ministry of Education and Scientific Research, Institute of Atomic Physics, Ministry of Research and Innovation and Institute of Atomic Physics and Universitatea Nationala de Stiinta si Tehnologie Politehnica Bucuresti, Romania;
Ministry of Education, Science, Research and Sport of the Slovak Republic, Slovakia;
National Research Foundation of South Africa, South Africa;
Swedish Research Council (VR) and Knut \& Alice Wallenberg Foundation (KAW), Sweden;
European Organization for Nuclear Research, Switzerland;
Suranaree University of Technology (SUT), National Science and Technology Development Agency (NSTDA) and National Science, Research and Innovation Fund (NSRF via PMU-B B05F650021), Thailand;
Turkish Energy, Nuclear and Mineral Research Agency (TENMAK), Turkey;
National Academy of  Sciences of Ukraine, Ukraine;
Science and Technology Facilities Council (STFC), United Kingdom;
National Science Foundation of the United States of America (NSF) and United States Department of Energy, Office of Nuclear Physics (DOE NP), United States of America.
In addition, individual groups or members have received support from:
European Research Council, Strong 2020 - Horizon 2020 (grant nos. 950692, 824093), European Union;
Academy of Finland (Center of Excellence in Quark Matter) (grant nos. 346327, 346328), Finland.

\end{acknowledgement}

\bibliographystyle{utphys}   
\bibliography{bibliography}

\newpage
\appendix

%
%

\section{The ALICE Collaboration}
\label{app:collab}
\begin{flushleft} 
\small

S.~Acharya\,\orcidlink{0000-0002-9213-5329}\,$^{\rm 127}$, 
D.~Adamov\'{a}\,\orcidlink{0000-0002-0504-7428}\,$^{\rm 87}$, 
A.~Adler$^{\rm 71}$, 
G.~Aglieri Rinella\,\orcidlink{0000-0002-9611-3696}\,$^{\rm 33}$, 
M.~Agnello\,\orcidlink{0000-0002-0760-5075}\,$^{\rm 30}$, 
N.~Agrawal\,\orcidlink{0000-0003-0348-9836}\,$^{\rm 52}$, 
Z.~Ahammed\,\orcidlink{0000-0001-5241-7412}\,$^{\rm 135}$, 
S.~Ahmad\,\orcidlink{0000-0003-0497-5705}\,$^{\rm 16}$, 
S.U.~Ahn\,\orcidlink{0000-0001-8847-489X}\,$^{\rm 72}$, 
I.~Ahuja\,\orcidlink{0000-0002-4417-1392}\,$^{\rm 38}$
% ...

\section*{Affiliation Notes}

$^{\rm I}$ Deceased\\
$^{\rm II}$ Also at: Max-Planck-Institut fur Physik, Munich, Germany\\
$^{\rm III}$ Also at: Italian National Agency for New Technologies, Energy and Sustainable Economic Development (ENEA), Bologna, Italy\\
$^{\rm IV}$ Also at: Dipartimento DET del Politecnico di Torino, Turin, Italy\\
$^{\rm V}$ Also at: Department of Applied Physics, Aligarh Muslim University, Aligarh, India\\
$^{\rm VI}$ Also at: Institute of Theoretical Physics, University of Wroclaw, Poland\\
$^{\rm VII}$ Also at: An institution covered by a cooperation agreement with CERN\\

\section*{Collaboration Institutes}

$^{1}$ A.I. Alikhanyan National Science Laboratory (Yerevan Physics Institute) Foundation, Yerevan, Armenia\\
$^{2}$ AGH University of Krakow, Cracow, Poland\\
$^{3}$ Bogolyubov Institute for Theoretical Physics, National Academy of Sciences of Ukraine, Kiev, Ukraine\\
$^{4}$ Bose Institute, Department of Physics  and Centre for Astroparticle Physics and Space Science (CAPSS), Kolkata, India\\
$^{5}$ California Polytechnic State University, San Luis Obispo, California, United States\\
$^{6}$ Central China Normal University, Wuhan, China\\
$^{7}$ Centro de Aplicaciones Tecnol\'{o}gicas y Desarrollo Nuclear (CEADEN), Havana, Cuba\\
$^{8}$ Centro de Investigaci\'{o}n y de Estudios Avanzados (CINVESTAV), Mexico City and M\'{e}rida, Mexico\\
$^{9}$ Chicago State University, Chicago, Illinois, United States\\
$^{10}$ China Institute of Atomic Energy, Beijing, China\\
$^{11}$ China University of Geosciences, Wuhan, China\\
$^{12}$ Chungbuk National University, Cheongju, Republic of Korea\\
$^{13}$ Comenius University Bratislava, Faculty of Mathematics, Physics and Informatics, Bratislava, Slovak Republic\\
$^{14}$ COMSATS University Islamabad, Islamabad, Pakistan\\
$^{15}$ Creighton University, Omaha, Nebraska, United States\\
$^{16}$ Department of Physics, Aligarh Muslim University, Aligarh, India\\
$^{17}$ Department of Physics, Pusan National University, Pusan, Republic of Korea\\
$^{18}$ Department of Physics, Sejong University, Seoul, Republic of Korea\\
$^{19}$ Department of Physics, University of California, Berkeley, California, United States\\
$^{20}$ Department of Physics, University of Oslo, Oslo, Norway\\
$^{21}$ Department of Physics and Technology, University of Bergen, Bergen, Norway\\
$^{22}$ Dipartimento di Fisica, Universit\`{a} di Pavia, Pavia, Italy\\
$^{23}$ Dipartimento di Fisica dell'Universit\`{a} and Sezione INFN, Cagliari, Italy\\
$^{24}$ Dipartimento di Fisica dell'Universit\`{a} and Sezione INFN, Trieste, Italy\\
$^{25}$ Dipartimento di Fisica dell'Universit\`{a} and Sezione INFN, Turin, Italy\\
$^{26}$ Dipartimento di Fisica e Astronomia dell'Universit\`{a} and Sezione INFN, Bologna, Italy\\
$^{27}$ Dipartimento di Fisica e Astronomia dell'Universit\`{a} and Sezione INFN, Catania, Italy\\
$^{28}$ Dipartimento di Fisica e Astronomia dell'Universit\`{a} and Sezione INFN, Padova, Italy\\
$^{29}$ Dipartimento di Fisica `E.R.~Caianiello' dell'Universit\`{a} and Gruppo Collegato INFN, Salerno, Italy\\
$^{30}$ Dipartimento DISAT del Politecnico and Sezione INFN, Turin, Italy\\
$^{31}$ Dipartimento di Scienze MIFT, Universit\`{a} di Messina, Messina, Italy\\
$^{32}$ Dipartimento Interateneo di Fisica `M.~Merlin' and Sezione INFN, Bari, Italy\\
$^{33}$ European Organization for Nuclear Research (CERN), Geneva, Switzerland\\
$^{34}$ Faculty of Electrical Engineering, Mechanical Engineering and Naval Architecture, University of Split, Split, Croatia\\
$^{35}$ Faculty of Engineering and Science, Western Norway University of Applied Sciences, Bergen, Norway\\
$^{36}$ Faculty of Nuclear Sciences and Physical Engineering, Czech Technical University in Prague, Prague, Czech Republic\\
$^{37}$ Faculty of Physics, Sofia University, Sofia, Bulgaria\\
$^{38}$ Faculty of Science, P.J.~\v{S}af\'{a}rik University, Ko\v{s}ice, Slovak Republic\\
$^{39}$ Frankfurt Institute for Advanced Studies, Johann Wolfgang Goethe-Universit\"{a}t Frankfurt, Frankfurt, Germany\\
$^{40}$ Fudan University, Shanghai, China\\
$^{41}$ Gangneung-Wonju National University, Gangneung, Republic of Korea\\
$^{42}$ Gauhati University, Department of Physics, Guwahati, India\\
$^{43}$ Helmholtz-Institut f\"{u}r Strahlen- und Kernphysik, Rheinische Friedrich-Wilhelms-Universit\"{a}t Bonn, Bonn, Germany\\
$^{44}$ Helsinki Institute of Physics (HIP), Helsinki, Finland\\
$^{45}$ High Energy Physics Group,  Universidad Aut\'{o}noma de Puebla, Puebla, Mexico\\
$^{46}$ Horia Hulubei National Institute of Physics and Nuclear Engineering, Bucharest, Romania\\
$^{47}$ HUN-REN Wigner Research Centre for Physics, Budapest, Hungary\\
$^{48}$ Indian Institute of Technology Bombay (IIT), Mumbai, India\\
$^{49}$ Indian Institute of Technology Indore, Indore, India\\
$^{50}$ INFN, Laboratori Nazionali di Frascati, Frascati, Italy\\
$^{51}$ INFN, Sezione di Bari, Bari, Italy\\
$^{52}$ INFN, Sezione di Bologna, Bologna, Italy\\
$^{53}$ INFN, Sezione di Cagliari, Cagliari, Italy\\
$^{54}$ INFN, Sezione di Catania, Catania, Italy\\
$^{55}$ INFN, Sezione di Padova, Padova, Italy\\
$^{56}$ INFN, Sezione di Pavia, Pavia, Italy\\
$^{57}$ INFN, Sezione di Torino, Turin, Italy\\
$^{58}$ INFN, Sezione di Trieste, Trieste, Italy\\
$^{59}$ Inha University, Incheon, Republic of Korea\\
$^{60}$ Institute for Gravitational and Subatomic Physics (GRASP), Utrecht University/Nikhef, Utrecht, Netherlands\\
$^{61}$ Institute of Experimental Physics, Slovak Academy of Sciences, Ko\v{s}ice, Slovak Republic\\
$^{62}$ Institute of Physics, Homi Bhabha National Institute, Bhubaneswar, India\\
$^{63}$ Institute of Physics of the Czech Academy of Sciences, Prague, Czech Republic\\
$^{64}$ Institute of Space Science (ISS), Bucharest, Romania\\
$^{65}$ Institut f\"{u}r Kernphysik, Johann Wolfgang Goethe-Universit\"{a}t Frankfurt, Frankfurt, Germany\\
$^{66}$ Instituto de Ciencias Nucleares, Universidad Nacional Aut\'{o}noma de M\'{e}xico, Mexico City, Mexico\\
$^{67}$ Instituto de F\'{i}sica, Universidade Federal do Rio Grande do Sul (UFRGS), Porto Alegre, Brazil\\
$^{68}$ Instituto de F\'{\i}sica, Universidad Nacional Aut\'{o}noma de M\'{e}xico, Mexico City, Mexico\\
$^{69}$ iThemba LABS, National Research Foundation, Somerset West, South Africa\\
$^{70}$ Jeonbuk National University, Jeonju, Republic of Korea\\
$^{71}$ Johann-Wolfgang-Goethe Universit\"{a}t Frankfurt Institut f\"{u}r Informatik, Fachbereich Informatik und Mathematik, Frankfurt, Germany\\
$^{72}$ Korea Institute of Science and Technology Information, Daejeon, Republic of Korea\\
$^{73}$ KTO Karatay University, Konya, Turkey\\
$^{74}$ Laboratoire de Physique Subatomique et de Cosmologie, Universit\'{e} Grenoble-Alpes, CNRS-IN2P3, Grenoble, France\\
$^{75}$ Lawrence Berkeley National Laboratory, Berkeley, California, United States\\
$^{76}$ Lund University Department of Physics, Division of Particle Physics, Lund, Sweden\\
$^{77}$ Nagasaki Institute of Applied Science, Nagasaki, Japan\\
$^{78}$ Nara Women{'}s University (NWU), Nara, Japan\\
$^{79}$ National and Kapodistrian University of Athens, School of Science, Department of Physics , Athens, Greece\\
$^{80}$ National Centre for Nuclear Research, Warsaw, Poland\\
$^{81}$ National Institute of Science Education and Research, Homi Bhabha National Institute, Jatni, India\\
$^{82}$ National Nuclear Research Center, Baku, Azerbaijan\\
$^{83}$ National Research and Innovation Agency - BRIN, Jakarta, Indonesia\\
$^{84}$ Niels Bohr Institute, University of Copenhagen, Copenhagen, Denmark\\
$^{85}$ Nikhef, National institute for subatomic physics, Amsterdam, Netherlands\\
$^{86}$ Nuclear Physics Group, STFC Daresbury Laboratory, Daresbury, United Kingdom\\
$^{87}$ Nuclear Physics Institute of the Czech Academy of Sciences, Husinec-\v{R}e\v{z}, Czech Republic\\
$^{88}$ Oak Ridge National Laboratory, Oak Ridge, Tennessee, United States\\
$^{89}$ Ohio State University, Columbus, Ohio, United States\\
$^{90}$ Physics department, Faculty of science, University of Zagreb, Zagreb, Croatia\\
$^{91}$ Physics Department, Panjab University, Chandigarh, India\\
$^{92}$ Physics Department, University of Jammu, Jammu, India\\
$^{93}$ Physics Program and International Institute for Sustainability with Knotted Chiral Meta Matter (SKCM2), Hiroshima University, Hiroshima, Japan\\
$^{94}$ Physikalisches Institut, Eberhard-Karls-Universit\"{a}t T\"{u}bingen, T\"{u}bingen, Germany\\
$^{95}$ Physikalisches Institut, Ruprecht-Karls-Universit\"{a}t Heidelberg, Heidelberg, Germany\\
$^{96}$ Physik Department, Technische Universit\"{a}t M\"{u}nchen, Munich, Germany\\
$^{97}$ Politecnico di Bari and Sezione INFN, Bari, Italy\\
$^{98}$ Research Division and ExtreMe Matter Institute EMMI, GSI Helmholtzzentrum f\"ur Schwerionenforschung GmbH, Darmstadt, Germany\\
$^{99}$ Saga University, Saga, Japan\\
$^{100}$ Saha Institute of Nuclear Physics, Homi Bhabha National Institute, Kolkata, India\\
$^{101}$ School of Physics and Astronomy, University of Birmingham, Birmingham, United Kingdom\\
$^{102}$ Secci\'{o}n F\'{\i}sica, Departamento de Ciencias, Pontificia Universidad Cat\'{o}lica del Per\'{u}, Lima, Peru\\
$^{103}$ Stefan Meyer Institut f\"{u}r Subatomare Physik (SMI), Vienna, Austria\\
$^{104}$ SUBATECH, IMT Atlantique, Nantes Universit\'{e}, CNRS-IN2P3, Nantes, France\\
$^{105}$ Sungkyunkwan University, Suwon City, Republic of Korea\\
$^{106}$ Suranaree University of Technology, Nakhon Ratchasima, Thailand\\
$^{107}$ Technical University of Ko\v{s}ice, Ko\v{s}ice, Slovak Republic\\
$^{108}$ The Henryk Niewodniczanski Institute of Nuclear Physics, Polish Academy of Sciences, Cracow, Poland\\
$^{109}$ The University of Texas at Austin, Austin, Texas, United States\\
$^{110}$ Universidad Aut\'{o}noma de Sinaloa, Culiac\'{a}n, Mexico\\
$^{111}$ Universidade de S\~{a}o Paulo (USP), S\~{a}o Paulo, Brazil\\
$^{112}$ Universidade Estadual de Campinas (UNICAMP), Campinas, Brazil\\
$^{113}$ Universidade Federal do ABC, Santo Andre, Brazil\\
$^{114}$ Universitatea Nationala de Stiinta si Tehnologie Politehnica Bucuresti, Bucharest, Romania\\
$^{115}$ University of Cape Town, Cape Town, South Africa\\
$^{116}$ University of Houston, Houston, Texas, United States\\
$^{117}$ University of Jyv\"{a}skyl\"{a}, Jyv\"{a}skyl\"{a}, Finland\\
$^{118}$ University of Kansas, Lawrence, Kansas, United States\\
$^{119}$ University of Liverpool, Liverpool, United Kingdom\\
$^{120}$ University of Science and Technology of China, Hefei, China\\
$^{121}$ University of South-Eastern Norway, Kongsberg, Norway\\
$^{122}$ University of Tennessee, Knoxville, Tennessee, United States\\
$^{123}$ University of the Witwatersrand, Johannesburg, South Africa\\
$^{124}$ University of Tokyo, Tokyo, Japan\\
$^{125}$ University of Tsukuba, Tsukuba, Japan\\
$^{126}$ Universit\"{a}t M\"{u}nster, Institut f\"{u}r Kernphysik, M\"{u}nster, Germany\\
$^{127}$ Universit\'{e} Clermont Auvergne, CNRS/IN2P3, LPC, Clermont-Ferrand, France\\
$^{128}$ Universit\'{e} de Lyon, CNRS/IN2P3, Institut de Physique des 2 Infinis de Lyon, Lyon, France\\
$^{129}$ Universit\'{e} de Strasbourg, CNRS, IPHC UMR 7178, F-67000 Strasbourg, France, Strasbourg, France\\
$^{130}$ Universit\'{e} Paris-Saclay, Centre d'Etudes de Saclay (CEA), IRFU, D\'{e}partment de Physique Nucl\'{e}aire (DPhN), Saclay, France\\
$^{131}$ Universit\'{e}  Paris-Saclay, CNRS/IN2P3, IJCLab, Orsay, France\\
$^{132}$ Universit\`{a} degli Studi di Foggia, Foggia, Italy\\
$^{133}$ Universit\`{a} del Piemonte Orientale, Vercelli, Italy\\
$^{134}$ Universit\`{a} di Brescia, Brescia, Italy\\
$^{135}$ Variable Energy Cyclotron Centre, Homi Bhabha National Institute, Kolkata, India\\
$^{136}$ Warsaw University of Technology, Warsaw, Poland\\
$^{137}$ Wayne State University, Detroit, Michigan, United States\\
$^{138}$ Yale University, New Haven, Connecticut, United States\\
$^{139}$ Yonsei University, Seoul, Republic of Korea\\
$^{140}$  Zentrum  f\"{u}r Technologie und Transfer (ZTT), Worms, Germany\\
$^{141}$ Affiliated with an institute covered by a cooperation agreement with CERN\\
$^{142}$ Affiliated with an international laboratory covered by a cooperation agreement with CERN.\\

\end{flushleft} 
  
\end{document}